\documentclass{article}
\usepackage{spconf}
\usepackage{url}
\usepackage{graphicx}
\usepackage{epstopdf}
\usepackage{algorithm}
\usepackage{algorithmicx}
\usepackage{algpseudocode}
\usepackage{cite}
\usepackage{color}
\usepackage{amsmath,amssymb,amsfonts,amsthm}
\usepackage{bm}
\usepackage{setspace}
\usepackage{multirow}
\usepackage{mathtools}
\usepackage{wrapfig}
\usepackage{subcaption}

\newtheorem{thm} {Theorem}

\newcommand{\ra}{\rightarrow}

\newcommand{\Lb}{\left[}
\newcommand{\Rb}{\right]}
\newcommand{\lb}{\left(}
\newcommand{\rb}{\right)}

\newcommand{\diag}{\textnormal{diag}}

\newcommand{\nmin}{{n_{\min}}}
\newcommand{\nmax}{{n_{\max}}}

\newcommand{\hatm}{\widehat{m}}

\newcommand{\hn}{\widehat{n}}

\newcommand{\htLB}{\widehat{t}_{\text{LB}}}

\newcommand{\tUB}{t_{\text{UB}}}
\newcommand{\tLB}{t_{\text{LB}}}

\newcommand{\hatt}{\widehat{t}}

\newcommand{\hp}{\widehat{p}}
\newcommand{\hpij}{\widehat{p}_{ij}}

\newcommand{\bone}{\mathbf{1}}
\newcommand{\bzero}{\mathbf{0}}

\newcommand{\bx}{\mathbf{x}}
\newcommand{\by}{\mathbf{y}}

\newcommand{\bv}{\mathbf{v}}

\newcommand{\bu}{\mathbf{u}}

\newcommand{\hL}{\widehat{\mathbf{L}}}

\newcommand{\Wbar}{\overline{W}}
\newcommand{\Wbarij}{\overline{W}_{ij}}

\newcommand{\hWbar}{\widehat{\Wbar}}
\newcommand{\hWbarij}{\widehat{\Wbar}_{ij}}

\newcommand{\SK}{S_{2:K}}

\newcommand{\bY}{\mathbf{Y}}

\newcommand{\btL}{\widetilde{\mathbf{L}}}

\newcommand{\btY}{\widetilde{\mathbf{Y}}}

\newcommand{\bA}{\mathbf{A}}

\newcommand{\bV}{\mathbf{V}}
\newcommand{\bW}{\mathbf{W}}

\newcommand{\bM}{\mathbf{M}}

\newcommand{\bL}{\mathbf{L}}

\newcommand{\bC}{\mathbf{C}}

\newcommand{\bS}{\mathbf{S}}

\newcommand{\hG}{\widehat{G}}

\begin{document}
\title{AMOS: An Automated Model Order Selection Algorithm for \\Spectral Graph Clustering}

\name{Pin-Yu Chen
	\thanks{This work was partially supported by Army Research Office grant W911NF-15-1-0479 and the Consortium for Verification
		Technology under Department of Energy National Nuclear Security Administration	award number DE-NA0002534.}
	\qquad Thibaut Gensollen
	 \qquad Alfred O. Hero III,~Fellow,~IEEE	
	}
\address{Department of Electrical Engineering and Computer Science, University of Michigan, Ann Arbor, USA \\ \{pinyu, thibautg, hero\}@umich.edu 
}

\ninept 
\maketitle
\begin{abstract}
One of the longstanding problems in spectral graph clustering (SGC) is the so-called model order selection problem: automated selection of the correct number of clusters. This is equivalent to the problem of finding the number of connected components or communities in an undirected graph. In this paper, we propose AMOS, an automated model order selection algorithm for SGC. Based on a recent analysis of clustering reliability for SGC under the random interconnection model, AMOS works by incrementally increasing the number of clusters, estimating the quality of identified clusters, and providing a series of clustering reliability tests. Consequently, AMOS outputs clusters of minimal model order with statistical clustering reliability guarantees. Comparing to three other automated graph clustering methods on real-world datasets, AMOS shows superior performance in terms of multiple external and internal clustering metrics.
\end{abstract}

\section{Introduction}

Undirected graphs are widely used for network data analysis, where nodes represent entities or data samples, and the existence and strength of edges represent relations or affinity between nodes.
The goal of graph clustering is to group the nodes into clusters of high similarity. Applications of graph clustering, also known as community detection \cite{White05,CPY14spectral}, include but are not limited to graph signal processing  \cite{Bertrand13Mag,Shuman13,Miller13detection,Dong2012clustering,Oselio14,Xu14,SihengChen15signalrecovery,Sandryhaila14,Wang15localset}, 
multivariate data clustering \cite{ng2002spectral,zelnik2004self,Luxburg07}, image segmentation \cite{Shi00,yu2002concurrent}, and network vulnerability assessment \cite{CPY14ComMag}. 

Spectral clustering \cite{ng2002spectral,zelnik2004self,Luxburg07} is a popular method for graph clustering, which we refer to as spectral graph clustering (SGC).
It works by transforming the graph adjacency matrix into a graph Laplacian matrix \cite{Merris94}, computing its eigendecomposition, and performing K-means clustering \cite{hartigan1979algorithm} on the eigenvectors to partition the nodes into clusters. 
Although heuristic methods have been proposed to automatically select the number of clusters \cite{Polito01grouping,ng2002spectral,zelnik2004self}, rigorous theoretical justifications on the selection of the number of eigenvectors for clustering are still lacking and little is known about the capabilities and limitations of spectral clustering on graphs.

Based on a recent development of clustering reliability analysis for SGC under the random interconnection model (RIM) \cite{CPY16AMOS}, we propose a novel automated model order selection (AMOS) algorithm for SGC.  AMOS works by incrementally increasing the number of clusters, estimating the quality of identified clusters, and providing a series of clustering reliability tests. Consequently, AMOS outputs clusters of minimal model order with statistical clustering reliability guarantees. Comparing the clustering performance  on real-world datasets, 
AMOS outperforms three other automated graph clustering methods in terms of multiple external and internal clustering metrics.

\section{Related Work}
 Most existing model selection algorithms specify an upper bound $K_{\max}$ on the number $K$ of clusters and then select $K$ based on optimizing some objective function, e.g., the goodness of fit of the $k$-cluster model for $k=2,\ldots, K_{\max}$. 
 In \cite{ng2002spectral}, the objective is to minimize the sum of cluster-wise Euclidean distances between each data point and the centroid obtained from K-means clustering. In \cite{Polito01grouping}, the objective is to maximize the gap between the $K$-th largest and the $(K+1)$-th largest eigenvalue. In \cite{zelnik2004self}, the authors propose to minimize an objective function that is associated with the cost of aligning the eigenvectors with a canonical coordinate system. In \cite{Newman06PNAS}, the authors propose to iteratively divide a cluster based on the leading eigenvector of the modularity matrix until no significant improvement in the modularity measure can be achieved. The Louvain method in \cite{blondel2008fast} uses a greedy algorithm for modularity maximization.
 In \cite{Krzakala2013,Saade2015spectral}, the authors propose to use the eigenvectors of the nonbacktracking matrix for graph clustering, where the number of clusters is determined by the number of real eigenvalues with magnitude larger than the square root of the largest eigenvalue.  
 The proposed AMOS algorithm not only automatically selects the number of clusters but also provides multi-stage statistical tests for evaluating clustering reliability of SGC.

\section{Theoretical Framework for AMOS}
\label{sec_AMOS_thm}
\subsection{Random interconnection model (RIM)}
\label{subsec_RIM}
Consider an undirected graph where its connectivity structure is represented by an $n \times n$ binary symmetric adjacency matrix $\bA$, where $n$ is the number of nodes in the graph. $[\bA]_{uv}=1$ if there exists an edge between the node pair ($u,v$), and otherwise $[\bA]_{uv}=0$. An unweighted undirected graph is completely specified by its adjacency matrix $\bA$, while a weighted undirected graph is specified by a nonnegative matrix $\bW$, where nonzero entries denote the edge weights.

Assume there are $K$ clusters in the graph and denote the size of cluster $k$ by $n_k$. The size of the largest and smallest cluster is denoted by $\nmax$ and $\nmin$, respectively.
Let $\bA_k$ denote the $n_k \times n_k$ adjacency matrix representing the internal edge connections in cluster $k$ and let $\bC_{ij}$ ($i,j \in \{ 1,2,\ldots,K\}$,~$i \neq j$) be an $n_i \times n_j$ matrix representing the adjacency matrix of inter-cluster edge connections between the cluster pair ($i,j$).
The matrix $\bA_k$ is symmetric and $\bC_{ij}=\bC_{ji}^T$ for all $i\neq j$.

The random interconnection model (RIM) \cite{CPY16AMOS}
assumes that: (1) the adjacency matrix $\bA_k$ is associated with a connected graph of $n_k$ nodes but is otherwise arbitrary; (2) the $K(K-1)/2$ matrices $\{\bC_{ij}\}_{i>j}$ are random mutually independent, and each $\mathbf C_{ij}$ has i.i.d. Bernoulli distributed entries with Bernoulli parameter $p_{ij} \in [0,1]$. (3) For undirected weighted graphs the edge weight of each inter-cluster edge between clusters $i$ and $j$ is independently drawn from a common nonnegative distribution with mean $\Wbarij$ and bounded fourth moment. In particular,
We call this model a \textit{homogeneous} RIM when all random interconnections have equal probability and mean edge weight, i.e., $p_{ij}=p$ and $\Wbarij=\Wbar$ for all $i \neq j$.  Otherwise, the model is called an \textit{inhomogeneous} RIM.


\subsection{Spectral graph clustering (SGC)}
The graph Laplacian matrix of the entire graph is defined as $\bL=\bS-\bW$, where $\bS=\diag(\bW\bone_n)$ is a diagonal matrix and $\bone_n (\bzero_n)$ is the $n \times 1$ column vector of ones (zeros). Similarly, the graph Laplacian matrix accounting for the within-cluster edges of cluster $k$ is denoted by $\bL_k$. We also denote the $i$-th smallest eigenvalue of $\bL$ by $\lambda_i(\bL)$ and define the partial eigenvalue sum $\SK(\bL)=\sum_{i=2}^{K} \lambda_i(\bL)$.
To partition the nodes in the graph into $K$ ($K \geq 2$) clusters, spectral clustering \cite{Luxburg07} uses the $K$ eigenvectors $\{\bu_k\}_{k=1}^K$ associated with the $K$ smallest eigenvalues of $\bL$. Each node can be viewed as a $K$-dimensional vector in the subspace spanned by these eigenvectors. K-means clustering \cite{hartigan1979algorithm} is then implemented on the $K$-dimensional vectors to group the nodes into $K$ clusters.
 
Throughout this paper we assume the graph is connected, otherwise the connected components can be easily found and
the proposed algorithm can be applied to each connected component separately. If the graph is connected, by the definition of the graph Laplacian matrix $\bL$, the smallest eigenvector $\bu_1$ is a constant vector and $\lambda_i(\bL)>0$~$\forall~i\geq 2$. As a result, for connected undirected graphs, it suffices to use the $K-1$ eigenvectors  $\{\bu_k\}_{k=2}^K$ of $\bL$ for SGC. In particular, these $K-1$ eigenvectors are  represented by the columns of the eigenvector matrix $\bY=[\bu_2,\bu_3,\ldots,\bu_K] \in \mathbb{R}^{n \times (K-1)}$.

\subsection{Phase transitions under homogeneous RIM}
	Let $\bY=[\bY_1^T,\bY_2^T,\ldots,\bY_K^T]^T$ be the cluster partitioned eigenvector matrix associated with  $\bL$ for SGC, 	where  $\bY_k \in \mathbb{R}^{n_k \times (K-1)}$ with its rows indexing the nodes in cluster $k$.  Under the homogeneous RIM, let $t=p \cdot \Wbar$ be the inter-cluster edge connectivity parameter. Fixing the within-cluster edge connections and  varying $t$, Theorem \ref{thm_spec} below shows that there exists a critical value $t^*$ that separates the behavior of $\bY$ for the cases of $t< t^*$ and $t>t^*$.

\begin{thm}
	\label{thm_spec}
		Under the homogeneous RIM with parameter $t=p \cdot \Wbar$,	
	there exists a critical value $t^*$ such that the following holds almost surely as $n_k \ra \infty$~$\forall~k \in \{1,2,\ldots,K\}$ and $\frac{\nmin}{\nmax} \ra c >0$: \\	
	\textnormal{(a)}~$\left\{	
	\begin{array}{ll}
	\textnormal{If~} t < t^*,~\bY_k = \bone_{n_k} \bone_{K-1}^T \bV_k\\
	~~~~~~~~~~~~~~~~~~~~~~~=\Lb v^k_1 \bone_{n_k},v^k_2 \bone_{n_k},\ldots,v^k_{K-1} \bone_{n_k} \Rb,~\forall~k; \\
	\textnormal{If~} t > t^*,~
	\bY_k^T\bone_{n_k} = \bzero_{K-1},~\forall~k; \\
	\textnormal{If~} t = t^*,~ \bY_k =\bone_{n_k} \bone_{K-1}^T \bV_k \textnormal{~or~} \bY_k^T\bone_{n_k} = \bzero_{K-1}, ~\forall~k,
	\end{array}
	\right.$ \\
	where $\bV_k=\diag(v^k_1, v^k_2,\ldots, v^k_{K-1}) \in \mathbb{R}^{(K-1) \times (K-1)}$.\\
		In particular, when $t < t^*$, $\bY$ has the following properties:\\
		\textnormal{(a-1)} The columns of $\bY_k$ are constant vectors. \\
		\textnormal{(a-2)} Each column of $\bY$ has at least two nonzero cluster-wise constant components, and these constants have alternating signs such that their weighted sum equals $0$ (i.e., $\sum_{k} n_k v^k_j = 0,~\forall~j \in\{1,2,\ldots,K-1\}$). \\
		\textnormal{(a-3)} No two columns of $\bY$ have the same sign on the cluster-wise nonzero components.	 \\	
	Furthermore, $t^*$ satisfies: \\
	\textnormal{(b)}~$\tLB \leq t^* \leq \tUB$, where \\
	$\tLB = \frac{\min_{k \in \{1,2,\ldots,K\}} \SK(\bL_k)}{(K-1)\nmax}$; ~
	$\tUB  = \frac{\min_{k \in \{1,2,\ldots,K\}} \SK(\bL_k)}{(K-1)\nmin}$.
 \\
	In particular, 	$\tLB=\tUB$ when $c=1$.	
	\end{thm}	
Theorem \ref{thm_spec}  (a) shows that there exists a critical value $t^*$ that separates the behavior of the rows of $Y$ into two regimes: (1) when $t<t^*$, based on conditions (a-1) to (a-3), the rows of each $\bY_k$ is identical and cluster-wise distinct  such that SGC can be successful. (2) when $t>t^*$, the row sum of each $\bY_k$ is zero, and the incoherence of the entries in $\bY_k$ make it impossible for SGC to separate the clusters. Theorem \ref{thm_spec}  (b)  provides closed-form upper and lower bounds on the critical value $t^*$, and these two bounds become tight when every cluster has identical size (i.e., $c=1$).

\subsection{Phase transitions under inhomogeneous RIM}
We can extend the phase transition analysis of the homogeneous RIM to the inhomogeneous RIM.  Let $\bY \in \mathbb{R}^{n \times (K-1)} $ be the eigenvector matrix of $\bL$ under the inhomogeneous RIM, and let $\btY \in \mathbb{R}^{n \times (K-1)}$ be the eigenvector matrix of the graph Laplacian $\btL$ of another random graph, independent of $\bL$, generated by a homogeneous RIM with cluster interconnectivity parameter $t$. 
We can specify the distance between the subspaces spanned by the columns of $\bY$ and $\btY$ by inspecting their principal angles \cite{Luxburg07}. Since $\bY$ and $\btY$ both have orthonormal columns, the vector $\bv$ of $K-1$ principal angles between their column spaces is $\bv=[\cos^{-1}\sigma_1(\bY^T \btY),\ldots,\cos^{-1}\sigma_{K-1}(\bY^T \btY)]^T$, where $\sigma_k(\bM)$ is the $k$-th largest singular value of real rectangular matrix $\bM$.
Let $\mathbf{\Theta}(\bY,\btY)=\diag(\bv)$, and let $\sin\mathbf{\Theta}(\bY,\btY)$ be defined entrywise. When $t<t^*$, the following theorem provides an upper bound on the 
Frobenius norm of $\sin\mathbf{\Theta}(\bY,\btY)$, denoted by $\| \sin\mathbf{\Theta}(\bY,\btY) \|_F$.

\begin{thm}
	\label{thm_principal_angle}
	Under the inhomogeneous RIM with interconnection parameters $\{t_{ij}=p_{ij} \cdot \Wbarij\}$, let $t^*$ be the critical threshold value for the homogeneous RIM specified by Theorem \ref{thm_spec}, and define $\delta_{t,n}=\min\{t,|\lambda_{K+1}(\frac{\bL}{n})-t|\}$. 
	For a fixed $t$, if $t < t^*$ and $\delta_{t,n} \ra \delta_t > 0$ as $n_k \ra \infty$~$\forall~k \in \{1,2,\ldots,K\}$,
	the following statement holds almost surely as
	$n_k \ra \infty$~$\forall~k$ and $\frac{\nmin}{\nmax} \ra c >0$:
	\begin{align}
	\|\sin\mathbf{\Theta}(\bY,\btY)\|_F \leq \frac{\| \bL - \btL \|_F}{n \delta_t}. \nonumber
	\end{align}
	Furthermore, let $t_{\max}=\max_{i \neq j} t_{ij}$. If $t_{\max} < t^*$, then 
	\begin{align}
		\|\sin\mathbf{\Theta}(\bY,\btY)\|_F \leq \min_{t \leq t_{\max}} \frac{\| \bL - \btL \|_F}{n \delta_t}. \nonumber
	\end{align}

\end{thm} 
By Theorem \ref{thm_spec}, since under the homogeneous RIM the rows of  $\btY$ has cluster-wise separability when $t<t^*$, Theorem \ref{thm_principal_angle} shows that under the inhomogeneous RIM cluster-wise separability in $\bY$ can still be expected provided that the subspace distance $\|\sin\mathbf{\Theta}(\bY,\btY)\|_F $ is small and $t<t^*$.  Moreover, if $t_{\max} < t^*$, we can obtain a tighter upper bound on $\|\sin\mathbf{\Theta}(\bY,\btY)\|_F $. These two theorems serve as the cornerstone of the proposed AMOS algorithm, and the proofs are given in the extended version \cite{CPY16AMOS}.

\section{Automated Model Order Selection (AMOS) Algorithm for Spectral Graph Clustering}
\label{sec_ASGC}
\begin{figure}[t!]		
	\centering
	\includegraphics[scale=0.45]{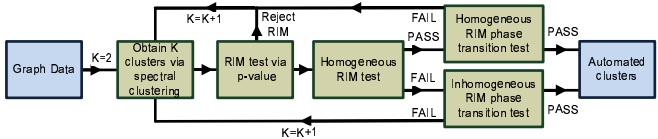}
		\vspace{-7mm}
	\caption{Flow diagram of the proposed automated model order selection (AMOS) algorithm for spectral graph cluster (SGC).}
	\label{Fig_automated_clustering}     
	\vspace{-4mm}
\end{figure}

Based on the theoretical framework in Sec. \ref{sec_AMOS_thm}, we propose an automated model order selection (AMOS) algorithm for automated cluster assignment for SGC. The flow diagram of AMOS is displayed in Fig. \ref{Fig_automated_clustering}, and the algorithm is summarized in Algorithm \ref{algo_automated_clustering}. The AMOS codes can be downloaded from  https://github.com/tgensol/AMOS.

AMOS works by iteratively increasing the number of clusters $K$ and performing multi-stage statistical clustering reliability tests until the identified clusters are deemed reliable.
The statistical tests in AMOS are implemented in two phases. The first phase is to test the RIM assumption based on the interconnectivity pattern of each cluster (Sec. \ref{subsec_RIM_test}), and the second phase is to test the homogeneity and variation of the interconnectivity parameter $p_{ij}$ for every cluster pair $i$ and $j$  in addition to making comparisons to the critical phase transition threshold (Sec. \ref{subsec_phase_transition_estimator}).  The proofs of the established statistical clustering reliability tests are given in the extended version \cite{CPY16AMOS}.

The input graph data of AMOS is a matrix representing a connected undirected weighted graph. For each iteration in $K$, SGC is implemented to produce $K$ clusters $\{\hG_k\}_{k=1}^K$, where  $\hG_k$ is the $k$-th identified cluster with number of nodes $\hn_k$ and number of edges $\hatm_k$. 

\begin{algorithm}[t]
	\caption{p-value computation of V-test for the RIM test}
	\label{algo_RIM_pvalue}
	\begin{algorithmic}
		\State \textbf{Input:} An $n_i \times n_j$ interconnection matrix $\widehat{\bC}_{ij}$
		\State \textbf{Output:} \text{p-value}$(i,j)$
		\State $\bx= \widehat{\bC}_{ij} \bone_{n_j}$~(\# of nonzero entries of each row in $\widehat{\bC}_{ij}$)
		\State $\by= n_j \bone_{n_i}-\bx$~(\# of zero entries of each row in $\widehat{\bC}_{ij}$)	
		\State $X=\bx^T \bx - \bx^T \bone_{n_i}$ and  $Y=\by^T \by - \by^T \bone_{n_i}$.
		\State $N=n_i n_j (n_j-1)$ and $V=\lb \sqrt{X} + \sqrt{Y} \rb^2$.
		\State Compute test statistic $Z=\frac{V-N}{\sqrt{2N}}$
		\State Compute \text{p-value}$(i,j)$$=2 \cdot \min \{ \Phi(Z),1-\Phi(Z) \}$
	\end{algorithmic}
\end{algorithm} 

\begin{algorithm}[t]
	\caption{Automated model order selection (AMOS) algorithm for spectral graph clustering (SGC)}
	\label{algo_automated_clustering}
	\begin{algorithmic}
		\State \textbf{Input:} a connected undirected weighted graph,  p-value significance level $\eta$, RIM confidence interval parameters $\alpha$, $\alpha^\prime$
		\State \textbf{Output:} number of clusters $K$ and identified clusters $\{\hG_k\}_{k=1}^K$
		\State Initialization: $K=2$. Flag $=1$.
		\While{Flag$=1$}		
		\State Obtain $K$ clusters $\{\hG_k\}_{k=1}^K$ via spectral clustering ($*$)
	    \State \# \emph{Local homogeneity testing} \# 
		\For{$i=1$ to $K$}
		\For{$j=i+1$ to $K$}
		\State Calculate p-value($i,j$) from Algorithm \ref{algo_RIM_pvalue}.
		\If{p-value($i,j$) $\leq \eta$}{~Reject RIM}
		\State Go back to ($*$) with $K=K+1$.
		\EndIf
		\EndFor
		\EndFor 
		\State Estimate $\hp$, $\hWbar$, $\{ \hp_{ij}\}$, $\{ \hWbarij\}$, and $\htLB$ specified in Sec. \ref{subsec_phase_transition_estimator}.
		\State \# \emph{Homogeneous RIM test} \#		
		\If{$\hp$ lies within the confidence interval in (\ref{eqn_spectral_multi_confidence_interval})}
		\State \# \emph{Homogeneous RIM phase transition test} \#
		\If{$\hp \cdot \hWbar$$  < \htLB$}				 
		Flag$=0$.
		\Else~ 
		Go back to ($*$) with $K=K+1$.		
		\EndIf    
		\ElsIf{$\hp$ does not lie within the confidence interval in (\ref{eqn_spectral_multi_confidence_interval})}
		\State \# \emph{Inhomogeneous RIM phase transition test} \#
		\If{$\prod_{i=1}^K \prod_{j=i+1}^K F_{ij}\lb \frac{\htLB}{\hWbarij},\hpij \rb \geq 1-\alpha^\prime$}		
		\State Flag$=0$.
		\Else
		~Go back to ($*$) with $K=K+1$.			
		\EndIf
		
		\EndIf
		\EndWhile
		\State Output  $K$ clusters $\{\hG_k\}_{k=1}^K$.		
	\end{algorithmic}
\end{algorithm}

\subsection{RIM test via p-value for local homogeneity testing}
\label{subsec_RIM_test}

Given clusters $\{\hG_k\}_{k=1}^K$ obtained from SGC with model order $K$, let $\widehat{\bC}_{ij}$ be the $\hn_i \times \hn_j$ interconnection matrix of between-cluster edges connecting clusters $i$ and $j$. The goal of local homogeneity testing is to compute a p-value to test the hypothesis that the identified clusters  satisfy the RIM. More specifically, 
we are testing the null hypothesis that \emph{$\widehat{\bC}_{ij}$ is a realization of a random matrix with i.i.d. Bernoulli entries} (RIM) and the alternative hypothesis that \emph{$\widehat{\bC}_{ij}$ is a realization of a random matrix with independent Bernoulli entries} (not RIM), for all $i \neq j$, $i>j$.
To compute a p-value for the RIM test we use the V-test \cite{potthoff1966testing} for homogeneity testing of the row sums of each interconnection matrix $\widehat{\bC}_{ij}$. Specifically, the V-test tests that the rows of $\widehat{\bC}_{ij}$ are all identically distributed.
For any $\widehat{\bC}_{ij}$ the test statistic $Z$ of the V-test converges to a standard normal distribution as $n_i,n_j \ra \infty$, and the p-value for the hypothesis that the row sums of $\widehat{\bC}_{ij}$ are i.i.d. is
\text{p-value}$(i,j)=2 \cdot \min \{ \Phi(Z),1-\Phi(Z) \}$, where $\Phi(\cdot)$ is the cumulative distribution function (cdf) of the standard normal distribution.
The proposed V-test procedure is summarized in Algorithm \ref{algo_RIM_pvalue}. The RIM test on $\widehat{\bC}_{ij}$ rejects the null hypothesis if \text{p-value}$(i,j) \leq \eta$, where $\eta$ is the desired single comparison significance level.  The AMOS algorithm won't proceed to the phase transition test stage (Sec. \ref{subsec_phase_transition_estimator}) unless every $\widehat{\bC}_{ij}$ passes the RIM test.

\begin{table}[t]
	\centering 
	\begin{tabular}{l|l|l|l}
		\hline
		Dataset                                                   & Node               & Edge              & 	\begin{tabular}[c]{@{}l@{}}	Ground truth                \end{tabular}                                                \\ \hline
		\begin{tabular}[c]{@{}l@{}}	IEEE reliability test \\   system (RTS) \cite{Grigg99}  \end{tabular}
		&  		\begin{tabular}[c]{@{}l@{}}		73 power   \\  	stations  \end{tabular}
  & 	\begin{tabular}[c]{@{}l@{}}	108 \\   power lines   \end{tabular}   & \begin{tabular}[c]{@{}l@{}}3 power\\  subsystems\end{tabular}    \\ \hline  
		\begin{tabular}[c]{@{}l@{}}		Hibernia Internet \\   backbone map \cite{Knight11}  \end{tabular}	                  
		& 55 cities          & 	\begin{tabular}[c]{@{}l@{}}	162   \\  connections  \end{tabular}   & \begin{tabular}[c]{@{}l@{}}American \&  \\ Europe cities
		\end{tabular}                                                        \\ \hline
		\begin{tabular}[c]{@{}l@{}}			Cogent Internet \\   	backbone map \cite{Knight11}     \end{tabular}			
		& 197 cities         & 	\begin{tabular}[c]{@{}l@{}}	243   \\  connections  \end{tabular}   & \begin{tabular}[c]{@{}l@{}} American \& \\  Europe cities \end{tabular}  \\ \hline
			\begin{tabular}[c]{@{}l@{}}			Minnesota road  \\   map \cite{MATLAB_BGL_2}   \end{tabular}
		
	                 & 	\begin{tabular}[c]{@{}l@{}}		2640  \\  intersections  \end{tabular}   & 3302 roads        & None                                                       \\ \hline
		Facebook \cite{mcauley2012learning}                
	                 & 			4039   users    & 
	                 			\begin{tabular}[c]{@{}l@{}}		    88234 \\       friendships     \end{tabular}	                 
	               &  None                 \\ \hline	                 
	\end{tabular}
	\caption{Summary of real-world datasets.}
	\label{table_data_SGC}
	\vspace{-3mm}
\end{table}

\subsection{Phase transition tests}
\label{subsec_phase_transition_estimator}
Once the identified clusters $\{\hG_k\}_{k=1}^K$ pass the RIM test, 
one can empirically determine the reliability of the clustering results using the phase transition analysis in Sec. \ref{sec_AMOS_thm}. AMOS first tests the assumption of homogeneous RIM, and performs the \textit{homogeneous RIM phase transition test} by comparing the empirical estimate $\hatt$ of the interconnectivity parameter $t$ with the empirical estimate $\htLB$ of the lower bound $\tLB$ on $t^*$ based on Theorem \ref{thm_spec}. If the test on the assumption of homogeneous RIM fails, AMOS then performs the \textit{inhomogeneous RIM phase transition test}  
by comparing the empirical estimate $\hatt_{\max}$ of $t_{\max}$ with  $\htLB$ based on Theorem \ref{thm_principal_angle}.

\noindent{ $\bullet$ \textbf{Homogeneous RIM test}}:
The homogeneous RIM test is summarized as follows. Given clusters $\{\hG_k\}_{k=1}^K$, we estimate the interconnectivity parameters $\{\hp_{ij}\}$ by $\hp_{ij}=\frac{\hatm_{{ij}}}{\hn_i \hn_j}$, where $\hatm_{ij}$ is the number of inter-cluster edges between clusters $i$ and $j$, and $\hp_{ij}$  is the maximum likelihood estimator (MLE) of $p_{ij}$. Under the homogeneous RIM, the estimate of the parameter $p$ is $\hp=\frac{2(m-\sum_{k=1}^{K}\hatm_k)}{n^2-\sum_{k=1}^{K} \hn_k^2}$, where $\hatm_k$ is the number of within-cluster edges of cluster $k$ and $m$ is the total number of edges in the graph.  A generalized log-likelihood ratio test (GLRT) is used to test the validity of the homogeneous RIM.  By the Wilk's theorem \cite{wilks1938large}, an asymptotic $100(1-\alpha) \%$ confidence interval for $p$ in an assumed homogeneous RIM is 
\begin{align}
\label{eqn_spectral_multi_confidence_interval}
&\Bigg\{ p:
\xi_{\binom{K}{2}-1,1-\frac{\alpha}{2}} \leq 
2\sum_{i=1}^K \sum_{j=i+1}^K  \mathbb{I}_{\{\hp_{ij} \in (0,1)\}} \Lb \hatm_{ij} \ln \hp_{ij} \right.   \\
&\left. \left. +(\hn_i \hn_j -\hatm_{ij} ) \ln (1-\hp_{ij}) \Rb -2 \lb m-\sum_{k=1}^{K}\hatm_{k} \rb  \ln p  \right. \nonumber \\
& 
- \Lb n^2-\sum_{k=1}^{K} \hn_k^2  
-2\lb m-\sum_{k=1}^{K}\hatm_{k} \rb \Rb  \ln (1-p)  \leq \xi_{\binom{K}{2}-1,\frac{\alpha}{2}} 
\Bigg\}, \nonumber
\end{align} 
where $\xi_{q,\alpha}$ is the upper $\alpha$-th quantile of the central chi-square distribution with degree of freedom $q$. The clusters pass the homogeneous RIM test if $\hp$ is within the confidence interval specified in (\ref{eqn_spectral_multi_confidence_interval}).

\noindent{ $\bullet$ \textbf{Homogeneous RIM phase transition test}}:
By Theorem \ref{thm_spec}, if the identified clusters follow the homogeneous RIM, then they are deemed reliable when $\hatt < \htLB$, where $\hatt=\hp \cdot \hWbar$, $\hWbar$ is the average of all between-cluster edge weights, and $\htLB = \frac{\min_{k \in \{1,2,\ldots,K\}} \SK( \hL_k)}{(K-1)\hn_{\max}}$.

\noindent{ $\bullet$ \textbf{Inhomogeneous RIM phase transition test}}:
If the clusters fail the homogeneous RIM test,  we then use the maximum of MLEs of $t_{ij}$'s, denoted by $\hatt_{\max}=\max_{i > j}\hatt_{ij}$, as a test statistic for testing the null hypothesis $H_0$: \emph{$\hatt_{\max} < \tLB$} against the alternative hypothesis $H_1$: \emph{$\hatt_{\max} \geq \tLB$}.
The test accepts $H_0$ if $\hatt_{\max}<\tLB$ and hence by Theorem \ref{thm_principal_angle} the identified clusters are deemed reliable.
Using the Anscombe transformation on the $\hpij$'s for variance stabilization \cite{anscombe1948transformation}, let $A_{ij}(x)=\sin^{-1} \sqrt{\frac{x+\frac{c^\prime}{\hn_i \hn_j}}{1+\frac{2c^\prime}{\hn_i \hn_j}}}$, where $c^\prime=\frac{3}{8}$.
Under the null hypothesis that $\hatt_{\max} < \tLB$, from \cite[Theorem 2.1]{chang2000generalized},
an asymptotic $100(1-\alpha^\prime)\%$ confidence interval for $\hatt_{\max}$ is $[0,\psi]$,
where  $\psi(\alpha^\prime,\{\hatt_{ij}\})$ is a function of the precision parameter $\alpha^\prime \in [0,1]$ and $\{\hatt_{ij}\}$. Furthermore, it can be shown that verifying $\psi < \htLB$ is equivalent to checking the condition
\begin{align}
\label{eqn_spectral_multi_confidence_interval_ingomogeneous_RIM}
\prod_{i=1}^K \prod_{j=i+1}^K F_{ij} \lb \frac{\htLB}{\hWbarij},\hpij \rb \geq 1-\alpha^\prime,
\end{align}
where $F_{ij}(x,\hpij)= \Phi \lb  \sqrt{4\hn_i \hn_j+2} \cdot \lb A_{ij}(x)- A_{ij}(\hpij)\rb \rb \cdot \mathbb{I}_{\{\hpij \in (0,1)\}}+ \mathbb{I}_{\{\hpij<x\}}\mathbb{I}_{\{\hpij \in \{0,1\}\}}$, and $\mathbb{I}$ is the indicator function.

\begin{table}[t]
	\centering	
	\begin{tabular}{lllllll}
		\hline
		\multicolumn{1}{c|}{Dataset}                                                 & \multicolumn{1}{c}{Method}                                                                                & \multicolumn{1}{c}{NMI}                                                                       & \multicolumn{1}{c}{RI}                                                                        & \multicolumn{1}{c}{F}                                                                 & \multicolumn{1}{c}{C}                                                               & \multicolumn{1}{c}{NC}                                                                        \\ \hline
		\multicolumn{1}{c|}{\begin{tabular}[c]{@{}c@{}}IEEE RTS\\ (3)\end{tabular}}       & \multicolumn{1}{c}{\begin{tabular}[c]{@{}c@{}}AMOS (3)\\ Louvain (6)\\ NB (3)\\ ST (2)\end{tabular}}      & \multicolumn{1}{c}{\begin{tabular}[c]{@{}c@{}} \textbf{.89}\\ .74\\ .75\\ .73\end{tabular}} & \multicolumn{1}{c}{\begin{tabular}[c]{@{}c@{}}\textbf{.96}\\ .83\\ .87\\ .78\end{tabular}} & \multicolumn{1}{c}{\begin{tabular}[c]{@{}c@{}}\textbf{.94}\\ .67\\ .81\\ .74\end{tabular}} & \multicolumn{1}{c}{\begin{tabular}[c]{@{}c@{}}.046\\ .144\\ .070\\ \textbf{.021}\end{tabular}} & \multicolumn{1}{c}{\begin{tabular}[c]{@{}c@{}}.068\\ .169\\ .100\\ \textbf{.041}\end{tabular}} \\ \hline
		\multicolumn{1}{c|}{\begin{tabular}[c]{@{}c@{}}Hibernia \\ (2)\end{tabular}} & \multicolumn{1}{c}{\begin{tabular}[c]{@{}c@{}}AMOS (2)\\ Louvain (6)\\ NB (2)\\ ST (2)\end{tabular}}      & \multicolumn{1}{c}{\begin{tabular}[c]{@{}c@{}}\textbf{1.0}\\ .27\\ .73\\ .87\end{tabular}} & \multicolumn{1}{c}{\begin{tabular}[c]{@{}c@{}}\textbf{1.0}\\ .51\\ .89\\ .96\end{tabular}} & \multicolumn{1}{c}{\begin{tabular}[c]{@{}c@{}}\textbf{1.0}\\ .32\\ .90\\ .96\end{tabular}} & \multicolumn{1}{c}{\begin{tabular}[c]{@{}c@{}}.030\\ .222\\ .027\\ \textbf{.028}\end{tabular}} & \multicolumn{1}{c}{\begin{tabular}[c]{@{}c@{}}.057\\ .263\\ .053\\ \textbf{.050}\end{tabular}} \\ \hline
		\multicolumn{1}{c|}{\begin{tabular}[c]{@{}c@{}}Cogent\\ (2)\end{tabular}}    & \multicolumn{1}{c}{\begin{tabular}[c]{@{}c@{}}AMOS (4)\\ Louvain (11)\\ NB (3)\\ ST (14)\end{tabular}}    & \multicolumn{1}{c}{\begin{tabular}[c]{@{}c@{}}\textbf{.42}\\ .24\\ .26\\ .34\end{tabular}} & \multicolumn{1}{c}{\begin{tabular}[c]{@{}c@{}}\textbf{.62}\\ .54\\ .54\\ .54\end{tabular}} & \multicolumn{1}{c}{\begin{tabular}[c]{@{}c@{}}.53\\ .25\\ \textbf{.57}\\ .28\end{tabular}} & \multicolumn{1}{c}{\begin{tabular}[c]{@{}c@{}}\textbf{.036}\\ .186\\ .073\\ .148\end{tabular}} & \multicolumn{1}{c}{\begin{tabular}[c]{@{}c@{}}\textbf{.048}\\ .204\\ .109\\ .164\end{tabular}} \\ \hline
		\multicolumn{1}{c|}{\begin{tabular}[c]{@{}c@{}}Minnesota\\ (-)\end{tabular}} & \multicolumn{1}{c}{\begin{tabular}[c]{@{}c@{}}AMOS (46)\\ Louvain (33)\\ NB (35)\\ ST (100)\end{tabular}} & \multicolumn{1}{c}{-}                                                                         & \multicolumn{1}{c}{-}                                                                         & \multicolumn{1}{c}{-}                                                                         & \multicolumn{1}{c}{\begin{tabular}[c]{@{}c@{}}\textbf{.074}\\ .290\\ .140\\ .119\end{tabular}} & \multicolumn{1}{c}{\begin{tabular}[c]{@{}c@{}}\textbf{.076}\\ .299\\ .144\\ .120\end{tabular}} \\ \hline
		\multicolumn{1}{c|}{\begin{tabular}[c]{@{}c@{}}Facebook\\ (-)\end{tabular}} & \multicolumn{1}{c}{\begin{tabular}[c]{@{}c@{}}AMOS (5)\\ Louvain (17)\\ NB (55)\\ ST (7)\end{tabular}} & \multicolumn{1}{c}{-}                                                                         & \multicolumn{1}{c}{-}                                                                         & \multicolumn{1}{c}{-}                                                                         & \multicolumn{1}{c}{\begin{tabular}[c]{@{}c@{}}\textbf{.004}\\ .076\\ .478\\ .006\end{tabular}} & \multicolumn{1}{c}{\begin{tabular}[c]{@{}c@{}}\textbf{.004}\\ .079\\ .486\\ .007\end{tabular}} \\ \hline		                                                                                         
	\end{tabular}
	\caption{Performance comparison of automated graph clustering algorithms. The number in the parenthesis of the Dataset (Method) column shows the number of ground-truth (identified) clusters. ``F'' (``C'') stands for F-measure (conductance). ``-'' means not available due to lack of ground-truth cluster information.
		For each metric, the best method is highlighted in bold face.  }
	\label{table_AMOS_performance}
	\vspace{-2mm}
\end{table}

\section{Experiments on Real-world Datasets}

We implement the proposed AMOS algorithm on the real-world network datasets in Table \ref{table_data_SGC}, and compare the clustering results with three other automated graph clustering methods, including the self-tuning  method (ST) \cite{zelnik2004self}, the nonbacktracking matrix method  (NB) \cite{Krzakala2013,Saade2015spectral}, and the Louvain method \cite{blondel2008fast}. For AMOS, we use the degree normalized adjacency matrix \cite{Luxburg07} as the input graph data, and set $\alpha=\alpha^\prime=0.05$ and $\eta=10^{-5}$.
For performance evaluation, multiple clustering metrics are computed for assessing the clustering quality. These metrics are normalized mutual information (NMI) \cite{zaki2014data}, Rand index (RI) \cite{zaki2014data}, F-measure (F) \cite{zaki2014data}, conductance (C) \cite{Shi00}, and normalized cut (NC) \cite{Shi00}. For NMI, RI, and F, higher value means better clustering performance, whereas for C and NC, lower value means better clustering performance.

Table \ref{table_AMOS_performance} summarizes the clustering performance of the datasets in Table \ref{table_data_SGC}.
For each dataset, AMOS has the most clustering metrics of best performance among these four methods, which demonstrates the robustness and reliability of AMOS. 
In particular, for the datasets with ground-truth cluster information such that the external clustering metrics NMI, RI, and F can be computed, AMOS shows significant improvement over other methods.
In addition, for clustering metrics over which AMOS does not prevail, its performance is comparable to the best method.

\section{Conclusion}
\label{sec_conclusion}

This paper presents an automated model order selection (AMOS) algorithm for spectral graph clustering (SGC). Stemming from the phase transition analysis on the clustering reliability of SGC under the random interconnection model, AMOS performs iterative SGC and multi-stage statistical tests such that  it automatically finds the minimal number of clusters with statistical clustering reliability guarantees. Experiments on real-world datasets show that AMOS outperforms other three automated graph clustering methods in terms of multiple external and internal clustering metrics.

\clearpage
\bibliographystyle{IEEEtran}
\bibliography{IEEEabrv20160824,CPY_ref_20160902}

\end{document}